# Conversational Swarm Intelligence amplifies the accuracy of networked groupwise deliberations

Louis Rosenberg
Unanimous AI
Pismo Beach, California
Louis@Unanimous.ai

Gregg Willcox
Unanimous AI
Seattle, Washington
Gregg@Unanimous.ai

Hans Schumann
Unanimous AI
San Francisco, California
Hans@Unanimous.ai

Ganesh Mani
Carnegie Mellon University
Pittsburgh, Pennsylvania
ganeshm@andrew.cmu.edu

*Abstract*— Conversational Swarm Intelligence (CSI) is a communication technology that enables large, networked groups (25 to 2500 people) to hold real-time conversational deliberations online. Modeled on the dynamics of biological swarms, CSI enables the reasoning benefits of small-groups with the collective intelligence benefits of large-groups. In this pilot study, groups of 25 to 30 participants were asked to select players for a weekly Fantasy Football contest over an 11-week period. As a baseline, participants filled out a survey to record their player selections. As an experimental method, participants engaged in a real-time text-chat deliberation using a CSI platform called Thinkscape to collaboratively select sets of players. The results show that the real-time conversational group using CSI outperformed 66% of survey participants, demonstrating significant amplification of intelligence versus the median individual (p=0.020). The CSI method also significantly outperformed the most popular choices from the survey (the Wisdom of Crowd, p<0.001). These results suggest that CSI is an effective technology for amplifying the intelligence of groups engaged in real-time large-scale conversational deliberation and may offer a path to collective superintelligence.

## I. Introduction.

Collective Intelligence (CI) refers to the phenomenon in which human groups can make predictions that exceed the capabilities of individual members [1]. Traditional methods involve capturing input from individual participants via surveys that are aggregated statistically. Commonly referred to as "Wisdom of Crowds" (WoC), this process often produces results that are more accurate than the median individual, thus exhibiting intelligence amplification. Still, such methods are generally limited to simple problems such as numerical estimations or multiple-choice selections. In the real world, problems often involve multiple interconnected factors that must be decided in parallel, each involving complex tradeoffs across competing factors. Traditional CI methods do not provide a pathway for solving complex open-ended problems while amplifying intelligence.

To address these limitations, a new CI technology called Conversational Swarm Intelligence (CSI) has been developed to enable large groups (25 to 2500 people) to engage in real-time deliberations while harnessing their collective intelligence. It's based on the biological principle of Swarm Intelligence but is applied to real-time human conversations using a novel application of Large Language Models (LLMs). To quantify the value of CSI, a complex forecasting task was chosen – selecting players for a weekly Fantasy Football contest. What makes the problem complex is that participants must "purchase" multiple players for their team while staying within a fixed budget. This requires a set of tradeoffs, weighing the value of spending budget on certain positions versus others. In a traditional WoC approach such as polls, surveys, or prediction markets, participants cannot collaboratively form a strategic approach. In this study we explore if CSI can solve this limitation, providing the intelligence amplification benefits of large groups while maintaining the deliberative benefits of thoughtful conversation. In this way, Conversational Swarm Intelligence may enable groups to leverage and amplify their collective intelligence in complex and open-ended problems.

## II. Swarm Intelligence (SI)

In the natural world, many species have evolved the ability to make groupwise decisions that are more accurate than the individuals could make on their own. Biologists refer to the phenomenon as *Swarm Intelligence* and it operates very differently from traditional human CI methods. Instead of aggregating estimates, biological groups form real-time systems in which participants engage in a multi-directional tug-of-war, pushing and pulling on the decision until a solution emerges that best represents their collective sentiments. Honeybees, for example, face a complex decision when selecting a new location for their hive [2,3]. They solve this by having hundreds of scout bees search a 20 to 30 square mile area, the scouts identifying potential locations and bringing the information back to the colony. Picking the best site is a multivariable problem that involves many competing constraints. For example, the bees need to select a site that is large enough to store the honey they need for winter, is near a good source of pollen, is insulated for winter, ventilated for summer, and is near a water source.

A human would have difficulty selecting the best option across the competing constraints and yet honeybees have been shown to collectively solve it. They do it by forming a real-time system in which scout bees express their preference through body vibrations known as a "waggle dance." Through a real-time "deliberation" among these competing vibration signals, a single decision emerges that maximizes support from scouts, and it is usually the best decision [3]. In this way, Swarm Intelligence is not an aggregation of data but a real-time negotiation that debates and converges.

Artificial Swarm Intelligence (ASI) is a new CI method that was developed in 2015 to enable networked human groups to form systems similar to bee swarms [4]. Because humans lack the ability to waggle dance like bees, artificial swarms were enabled using a unique graphical interface [5]. Each user was given the ability to control a graphical magnet with a mouse or touchscreen such that large groups of simultaneous users can engage in a tug-of-war to collectively guide a graphical pointer while AI algorithms process their dynamic behaviors to infer sentiment strengths [6-8]. This technique has been validated through numerous studies. For example, researchers at MIT and Unanimous AI showed that groups of financial traders, when deliberating in swarms, amplified forecasting accuracy by 36% (p<0.001) compared to survey aggregation [9]. Researchers at Stanford and Unanimous AI showed that groups of doctors working in swarms reduced diagnostic errors by 33% as compared to traditional methods [7, 10]. In addition, researchers at California Polytechnic (Cal Poly) showed that human teams could significantly amplify "social perceptiveness" when deliberating in real-time swarms [11, 12].

While Artificial Swarm Intelligence enables human groups to collectively rank, rate, or select among pre-defined sets of options [6, 8], such tasks are too narrow to solve complex open-ended problems. That's because participants cannot adequately express their insights on open-ended problems. To address this, a new technology was introduced in 2023 known as Conversational Swarm Intelligence that amplifies the intellect of large, networked groups through real-time conversational deliberation [13].

### III. CONVERSATIONAL SWARM INTELLIGENCE (CSI)

Conversational Swarm Intelligence was developed to enable large groups to engage in real-time deliberations and converge on solutions that amplify their collective intelligence [14]. This required novel technology because current communication tools cannot enable coherent real-time conversations among large, networked groups. For example, large groups can congregate in real-time chatrooms, but that does not yield coherent deliberation. That's because conversational quality degrades with group size [15]. When groups grow beyond 4 to 7 people, the conversational dynamics quickly degrade, providing less "airtime" per person, disrupting turn-taking, and reducing engagement. In fact, putting dozens of individuals in a single chatroom or videoconference would not yield an authentic conversation. Instead, it would devolve into a series of monologues.

To solve this, CSI technology takes inspiration from the behavior of fish schools [16]. Large schools with thousands of members can hold coherent "conversations," enabling rapid decisions without a central authority to mediate. Each fish communicates with others using a unique organ called a *lateral line* that detects subtle pressure changes caused by neighboring fish as they adjust direction and speed. The number of neighbors that each fish detects varies among species but is always a small subset of the population. This begs the question – *if each fish can only attend to a small subset of others, how does the full population hold a unified real-time conversation?* The secret is overlapping subsets. Because each fish reacts to a subset of fish that overlaps the subsets attended to by other fish, information can rapidly propagate across the full school. This enables a unified Swarm Intelligence to emerge that quickly converges on unified decisions [17, 18]. In fact, the communication is so fast across the population, biologists often refer to fish schools as Super Organisms because they behave like a singular unified entity that acts and reacts within its world.

CSI was designed with a similar structure, enabling large, networked groups to hold real-time conversations such that each participant can have a coherent and thoughtful discussion with a small subset of their neighbors while also allowing deliberative information to rapidly propagate across the full population. For example, a networked group of 400 users might be divided into 80 groups of 5 people. This is done by routing the members of each subgroup into their own unique chatroom or videoconference. That said, merely subdividing the group into 80 subgroups does not enable a conversational collective intelligence unless information can propagate. Researchers solved this through the novel use of AI agents powered by Large Language Models [13, 14, 16].

To emulate the overlap across subgroups, a novel AI-powered Conversational Agent is inserted into each of the parallel rooms and tasked with continuously observing the deliberative dialog within that unique room, distilling salient content in terms of the assertions made within that room and the unique arguments expressed in favor or against such assertions. It is also tasked with expressing a concise representation of assertions and arguments in neighboring rooms as first-person dialog. Of course, the same process is happening in all rooms simultaneously. In this way, the CSI structure can be modeled as each group being given an additional member that is an AI agent that expresses the insights that emerge in one group into neighboring groups. This enables information to smoothly propagate across the population. Using this architecture, 50, 500 or even 5,000 people could hold a real-time deliberation, sharing views and ideas, debating options, and converging on unified solutions that optimize overall support. An example CSI structure for roughly 100 members is shown below in Figure 1.

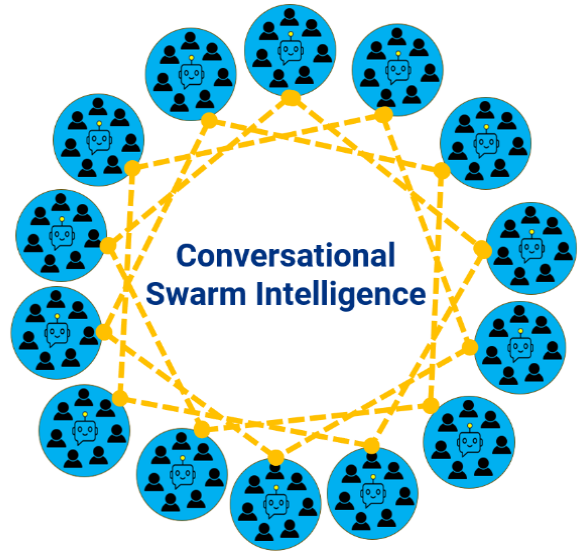

Fig. 1. Architecture for a Conversational Swarm with roughly 100 real-time members with artificial agents assigned to each group for passing content.

CSI has three significant advantages over prior CI tools and techniques. First, it enables groups to address complex problems and open-ended questions in which answer options are not defined up front, empowering participants to suggest and debate a wide range of alternatives as they naturally emerge and propagate. Second, it allows users to not only express which options they personally prefer but also argue for or against options with specific reasons, rationales, or justifications. And third, CSI is designed to assess and track sentiments in real-time by user or subgroup, assessing how new ideas or new arguments are received locally (within a subgroup) and globally as it propagates. *Do assertions take off and spread across the population, or do they hit resistance and fail to spread?* This information is tracked in real-time and is used to generate detailed analytics that reveal how and why the group reached the solutions it did.

In this way, CSI not only facilitates convergence of large groups on unified solutions, but captures the reasonings and rationales that underlie the process. In addition, CSI is designed to mitigate social influence bias because each member is only influenced by a small number of others in real-time, reducing the impact of early views and/or strong personalities on the full population [17]. In this way, CSI combines the intelligence amplification benefits of large groups with the deliberative reasoning of small groups. And because CSI is highly scalable, it could be used to connect thousands or even millions of real-time users. This means it could provide a pathway to collective superintelligence.

As for prior results, an early study conducted at Carnegie Mellon in 2023 tested real-time groups of 25 participants using the Thinkscape CSI platform and compared results to standard centralized chat. The participants in the CSI structure produced 30% more contributions (p<0.05) than those using standard chat and 7.2% less variance, indicating that users participated more evenly when using CSI [13].

In a larger CSI study, groups of 48 users were tasked with debating a current-event topic. The participants using CSI contributed 51% more content (p<0.001) compared to those using standard chat. In addition, CSI showed 37% less difference in contribution between the most vocal and least vocal users, indicating that CSI fosters more balanced deliberations. In addition, a large majority of the participants preferred the Thinkscape CSI platform over standard chat (p<0.05) and reported feeling more impactful (p<0.01) [14].

In an even larger study, a group of 80 participants was tested in Thinkscape to assess the ability of CSI to generate qualitative insights regarding a set of political candidates running for office in the United States in 2024. After only six minutes of chat-based deliberation, the group converged on a preferred candidate and surfaced over 200 reasons for supporting that candidate. The preferred solution converged with statistically significant sentiment within only six minutes (p<0.001) [16,17].

In a fourth study, 241 users engaged in a single largescale text-chat conversation using the Thinkscape platform. The group was asked to estimate the number of gumballs in a jar by viewing a photograph (see Figure 2). The CSI method partitioned the 241 participants into 47 subgroups of 5 or 6 members while AI agents passed conversational content around the network [29].

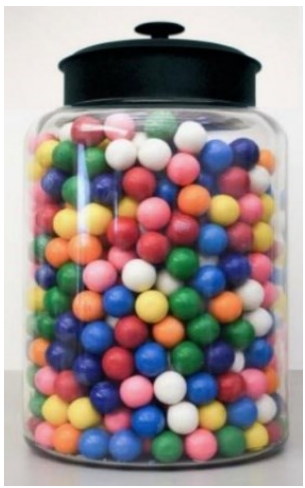

Fig 2. Image of gumballs in a jar.

The conversational estimates generated using Thinkscape were compared to a traditional survey-based aggregation across the same set of 241 users. In addition, GPT-4.0 was given the same photograph of the jar and asked to estimate the quantity of gumballs within. As shown in Figure 3 below, the CSI platform outperformed the average individual, the traditional wisdom of crowd, and the GPT-4.0 estimate. In fact, the CSI estimate had a 50% smaller error than the survey based WoC technique, a surprising result [30].

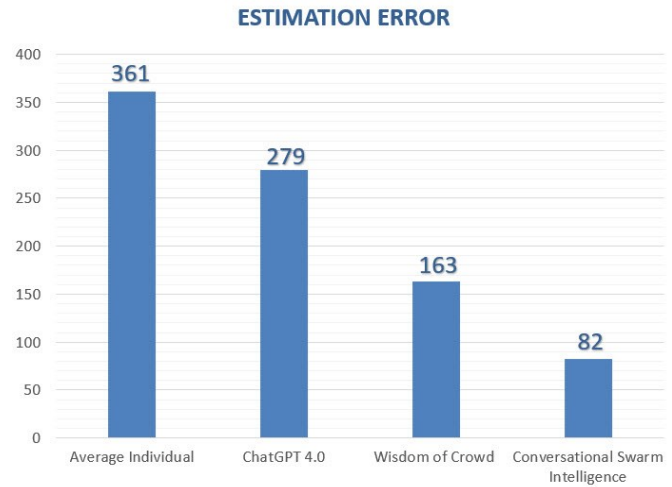

Fig 3. Comparison of estimation error across four test methods.

## IV. NFL Negotiation Study

In the current study, the Thinkscape platform was used to test the ability of distributed groups to solve a multi-faceted problem that requires strategy and planning for optimal performance – the selecting of players for a weekly Fantasy Football contest. This task is complex because participants must collaboratively select a set of players to "purchase" for their team while staying within a fixed budget. This requires tradeoffs, weighing the value of spending budget on certain positions versus others. For example, if the group decides to spend much of their budget on the best quarterback available, they may not have sufficient funds to get a solid running back or wide receiver. There is no perfect strategy, but effective collaboration means planning ahead and making difficult tradeoffs. An experimental question is whether such strategic planning is possible within the CSI architecture of networked subgroups working in parallel.

Sessions were conducted weekly over a period of 11 consecutive weeks during the 2023-2024 NFL season. Groups of 25 to 30 participants were engaged each week. All were self-identified NFL fans sourced from Amazon Mechanical Turk and all were familiar with fantasy sports. The individuals were first tasked with selecting players on their own using a standard survey with the goal of maximizing points scored. The group was then tasked with collaboratively selecting players via real-time conversation in Thinkscape, again with the goal of maximizing points scored. In each weekly session, we modeled the contest using data from DraftKings: simulating a budget and providing sets of available players to choose among.

To reduce the time required to generate a full fantasy roster, we pre-selected 4 of the 9 positions required and tasked the group with selecting the remaining five while staying within a fixed budget. After prefilling 4 of the 9 positions, the participants were given $32,500 remaining in the budget to select 5 positions. For each of the 5 positions (QB, RB, WR, TE, DST), 5 players of varying values were given as options, forcing the individuals to debate which positions were worth spending more on, and which are worth taking the less expensive option, selecting only one player per position. The order in which the positions were presented to the group was randomized each week.

When using Thinkscape, the group was given 4 minutes to select a player for each position, resulting in a little over a 20-minute session to select all five positions. The group was informed of their remaining budget before each question and never created a roster that went over budget.

To calculate a Wisdom of Crowd (WoC) roster, the pre-swarm individual surveys were aggregated and the most popular answer for each position was selected. In the two sessions when the WoC roster exceeded the budget, the position with the lowest plurality (the answer with the least support) was replaced with the next most popular answer for that position, and this procedure was repeated until budget constraints were met. This procedure was designed to capture the set of players that the group could best agree upon.

## V. RESULTS

Across the 11 sessions, the groupwise result generated in Thinkscape averaged 86.8 points per session in its selection of the 5 positions. As shown in Table 1, this result outperformed the median individual's score (77.2 points) when completing the task in isolation in the pre-swarm survey. A paired t-test by session indicates this difference is statistically significant (p=0.020). The Thinkscape group also significantly exceeded the Wisdom of the Crowd answers, which averaged 74.2 points across the 11 weeks (p<0.001).

Another way to measure performance is to determine the percentage of survey participants that Thinkscape outperformed. On average, Thinkscape exceeded the score of 66% of individually generated rosters. Using a bootstrap test and resampling 10,000 times over the observed participant performances, we can be 95% confident that Thinkscape would outperform between 61.4% to 69.7% of individuals on average over time.

| Aggregation Method | Average Points | Average Percentile Performance |
|---|---|---|
| Thinkscape (CSI) | 86.8 | 65.6% |
| Survey (WoC) | 74.2 | 40.9% |
| Survey (Median Individual) | 77.2 | 50.0% |

Table 1. Comparison of average performance in Fantasy Football

Figure 4 shows the distribution of score differences for each session between Thinkscape and the Median Individual: Thinkscape averaged 9.7 more points per session than the median individual and scored higher in 8 out of 11 sessions.

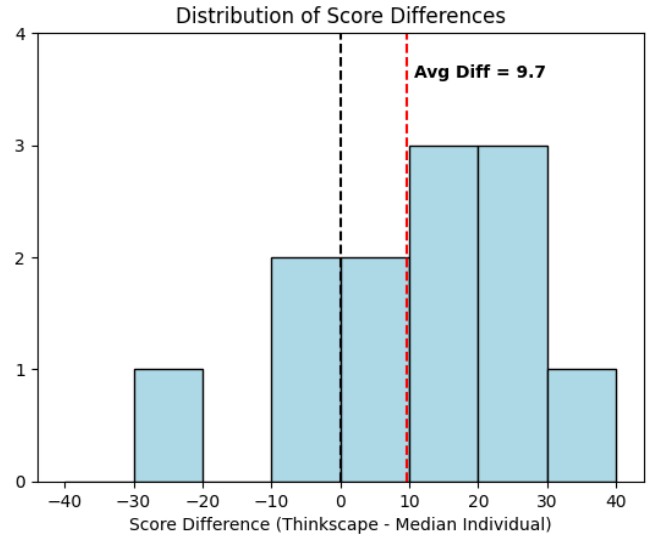

Fig 4. Distribution of score differences between Thinkscape and the Median Individual.

To measure the differences between the WoC and Thinkscape at a more granular level, we can compare the performance of each approach on each question. With 5 players selected in each of 11 sessions, there were 55 total player selections. As seen in Table 3, both approaches picked the same player for 40 of 55 questions (73%). On the 27% of remaining questions, Thinkscape picked a player that outperformed the WoC's choice 13 out of 15 times (24%) and only picked a worse-performing player 2 times (4%), resulting in a significantly higher pairwise performance (p=0.004). As a result, we can conclude that Thinkscape enabled this group to pick fantasy NFL players that were likely to outperform the WoC's picks with higher probability than random chance allows.

| Higher Scoring Method | Number of Players | Proportion of Player Selections |
|---|---|---|
| Thinkscape (CSI) performed best | 13 | 23.6% |
| Wisdom of Crowd (WoC) performed best | 2 | 3.6% |
| Same Score | 40 | 72.7% |

Table 2. Comparison of number of player selections where Thinkscape or Wisdom of Crowd scored more points.

## VI. CONCLUSIONS

We conducted a complex collaborative task using a new CI technology called Conversational Swarm Intelligence. It combines the methods inspired by the biological principle of Swarm Intelligence with a novel use of LLMs that enables large, distributed groups to hold coherent conversational deliberation and reach thoughtful and unified solutions. We performed an 11-week study with groups of 25 to 30 participants each week, the groups tasked with selecting five players for a Fantasy Football team that was scored using standard DraftKings contest rules. This task was chosen as it replicates common real-world business tasks in which groups must debate tradeoffs between options, weighing their relative costs against their potential outcomes.

Prior to running this study, it was unknown whether groups distributed across a real-time interconnected network of subgroups (each supported by an LLM-based AI agent) could successfully deliberate as a unified system and solve a complex problem that requires strategic planning over time. The results confirmed that the group using Thinkscape could (a) conversationally select a unified set of players (while staying within a fixed budget) by using the CSI networked architecture, (b) could significantly outperform the median individual in the group when scored on that team, and (c) could significantly outperform the teams collaboratively selected using a standard survey-based Wisdom of Crowd method. Specifically, it was found that when using the novel Thinkscape CSI platform to deliberate, the groups of participants significantly outperformed the median individual ($p=0.020$) and the most popular individual choices across individuals (WoC, $p<0.001$).

Overall, this study demonstrates that Conversational Swarm Intelligence or (CSI) is a viable method for amplifying the collective intelligence of networked human groups through a purely conversational means. Future work will test larger groups and more complex problems, for example groups in the many thousands of simultaneous users, with the goal of achieving collective superintelligence (CSi).

Significant opportunities exist to continue advancing conversational collective intelligence through CSI structures. For example, as LLMs and related technologies grow more capable, more diverse hybrid human-AI teams could be fielded in which subgroups include artificial agents that bring additional informational content and or analytical expertise into the mix. This could enable even broader ideas, perspectives, or arguments to propagate through networked deliberations, uncovering new problem-solving pathways.

Additionally, complex real-world problems can often benefit from both human and machine expertise. For example, developing policy to guide the use of emerging technologies requires understanding subtle cultural attitudes that humans can best provide along with real-time analytical capabilities that machines excel at. As such, integrating both human and machine voices in real-time CSI deliberations may lead to innovative solutions that would not otherwise be found.

ACKNOWLEDGMENT

The authors thank Chris Hornbostel and Patty Sullivan for their efforts recruiting participants and moderating sessions.